\documentclass[conference, 9pt]{IEEEtran}
\IEEEoverridecommandlockouts

\usepackage{cite}
\usepackage{amsmath,amssymb,amsfonts}
\usepackage{algorithmic}
\usepackage{graphicx}
\usepackage{textcomp}
\usepackage{xcolor}
\def\BibTeX{{\rm B\kern-.05em{\sc i\kern-.025em b}\kern-.08em
    T\kern-.1667em\lower.7ex\hbox{E}\kern-.125emX}}
\begin{document}
\topmargin=0mm

\title{ATFLRec: A Multimodal Recommender System with Audio-Text Fusion and Low-Rank Adaptation via Instruction-Tuned Large Language Model}

\author{\IEEEauthorblockN{Zezheng Qin}
\IEEEauthorblockA{\textit{Faculty of Computer Science and Information Technology} \\
\textit{Universiti Putra Malaysia}\\
Selangor, Malaysia \\
Zezheng.chin@gmail.com}
}

\maketitle
\vfill

\noindent This work has been submitted to the IEEE for possible publication. Copyright may be transferred without notice, after which this version may no longer be accessible.

\begin{abstract}
Recommender Systems (RS) play a pivotal role in boosting user satisfaction by providing personalized product suggestions in domains such as e-commerce and entertainment. This study examines the integration of multimodal data—text and audio—into large language models (LLMs) with the aim of enhancing recommendation performance. Traditional text and audio recommenders encounter limitations such as the cold-start problem, and recent advancements in LLMs, while promising, are computationally expensive. To address these issues, Low-Rank Adaptation (LoRA) is introduced, which enhances efficiency without compromising performance. The ATFLRec framework is proposed to integrate audio and text modalities into a multimodal recommendation system, utilizing various LoRA configurations and modality fusion techniques. Results indicate that ATFLRec outperforms baseline models, including traditional and graph neural network-based approaches, achieving higher AUC scores. Furthermore, separate fine-tuning of audio and text data with distinct LoRA modules yields optimal performance, with different pooling methods and Mel filter bank numbers significantly impacting performance. This research offers valuable insights into optimizing multimodal recommender systems and advancing the integration of diverse data modalities in LLMs.
\end{abstract}


\section{Introduction}
During activities such as listening to music, reading novels, and online shopping, Provider can leverage historical data and user preferences to recommend products that better align with user tastes, thereby increasing user satisfaction\cite{tongxiao2011value}. This is made possible with the assistance of recommender systems.

Recommender Systems (RS) are information filtering systems designed to predict and suggest items or content that users may find interesting, such as products, movies, music, or articles\cite{singh2021recommender}. These predictions are based on past user behaviors, preferences, or the behaviors of similar users. The primary goal of RS is to enhance user experience, increase engagement, and facilitate decision-making processes. This applies to various domains, including e-commerce, entertainment, and social media. Based on the primary data modalities of the recommended content, recommender systems are categorized into audio recommender systems, text recommender systems, multimodal recommender systems, and others\cite{deldjoo2021multimedia}.

Audio recommender systems are designed to recommend personalized audio content based on users' listening habits, preferences, and behaviors. Audio recommender systems find extensive application in music recommendations, audiobook recommendations and short video recommendations\cite{afchar2022explainability}. For short video recommendations, audio recommender systems utilize fewer training and deployment resources compared to directly using video data for recommendations\cite{deldjoo2021multimedia}.

Text recommender systems process and analyze text data generated from user interactions with the system, employing methods such as content-based filtering, collaborative filtering, and machine learning approaches, including GRU4Rec \cite{jannach2017recurrent} and Graph Neural Network-based recommendation systems \cite{wu2022graph}. However, these methods struggle with the cold-start problem\cite{silva2019pure}. With the rapid development of large language models (LLMs), LLMs have demonstrated outstanding capabilities in recommender systems\cite{zhao2024recommender}. Nonetheless, the substantial number of parameters in these LLM-based systems makes adapting the entire system to performing recommended work computationally impractical and costly. Low-Rank Adaptation (LoRA) \cite{hu2021lora}addresses these issues by modifying specific system parameters using low-rank matrices, showing great promise. This approach is memory-efficient during training and does not impact the runtime of recommender systems\cite{zhou2024survey}.

As a result, recommender systems leveraging the fine-tuning of large language models have gained popularity. These approaches utilize a few-shot training setup, selecting a limited number of samples from the training set for model training. While they effectively address the cold-start problem, they do not account for other data modalities present in recommender system data\cite{bao2023tallrec,cui2022m6}. In many cases, purely text-based interactions with LLMs may be limited, as they often fail to capture information that is difficult to convey through text alone. For instance, audio can encode a range of emotions in speech, while images can represent geometric shapes and object positions, which may be challenging to describe in text\cite{fathullah2024prompting}.

To address these issues, multimodal recommender models have emerged. Multimodal recommender systems combine multiple data modalities (such as text, images, audio, video, etc.) for content recommendation. They are more complex than traditional single-modal recommender systems because they need to simultaneously handle and integrate information from different modalities, providing more accurate and personalized recommendations\cite{truong2021multi}. Although multimodal data offers more comprehensive user and content information, making recommendations more personalized and precise, effectively integrating modalities and enhancing the recommendation performance of large language models remains a significant challenge\cite{deldjoo2021multimedia}. Additionally, as noted by \cite{fathullah2024prompting}, fine-tuning large language models with a combination of audio and text content can significantly improve their inference capabilities. However, this study only integrates text and audio information into a single LoRA module for fine-tuning, without exploring the potential benefits of separately fine-tuning audio and text modalities on the performance of large language models.

To address these challenges, this paper proposes the following research questions:
RQ1: How can audio data and text information be effectively integrated to achieve collaborative recommendation?
RQ2: How does the proposed method perform compared to other models?
RQ3: What are the differences in training recommender systems using various low-rank adaptation modules?
RQ4: What are the effects of using different modality aggregation methods on the performance of recommender systems?

Based on these questions and to better leverage the rich knowledge capabilities of large language models, this paper integrates audio modality content into large language models for joint recommendation. Additionally, the model attempts fine-tuning using various numbers and combinations of LoRA modules at a low parameter level. To the best of the author's knowledge, this is the first study to integrate audio and text data modalities while employing instruction-tuning of large language models for fine-tuning and joint recommendations.

The main contributions of this paper are as follows: 
1. Proposing a multimodal recommendation system that integrates audio modality content into large language models.
2. Exploring the impact of different LoRA modules at a low parameter level on large language models, providing empirical insights for fine-tuning multimodal models.
3. Investigating the effects of audio stacking pooling methods, multimodal data fusion pooling methods, and the number of filters on recommendation system performance when processing audio data.

\section{Methodology}

This work focuses on integrating text and audio modalities within large language model recommender systems, with particular emphasis on exploring methods to enhance the performance of multimodal recommender systems.

\subsection{Model Framework}

This section introduces the ATFLRec framework, aimed at facilitating the integration of audio and text data and the effective and efficient alignment of LLMs with recommendation tasks, especially in low GPU memory settings. The ATFLRec framework includes audio extraction and embedding, alignment of audio and text features, and the setup of the LoRA fine-tuning model.

The overall framework of ATFLRec is illustrated in Figure 1. Initially, user historical data is matched with item data to generate textual instructions, as shown in Table I. The text is then tokenized using the LLM's tokenizer and fed into the embedding module to obtain text features. Simultaneously, FBank feature extraction is applied to the audio of user-liked and target items, followed by deep feature processing. The audio features of item and target items are then fused and aligned dimensionally with the text features. Finally, both text and audio features are input into the multimodal large model recommender system, where fine-tuning is performed using LoRA modules based on the labels.
\begin{figure}[h]
    \centering
    \includegraphics[width=\linewidth]{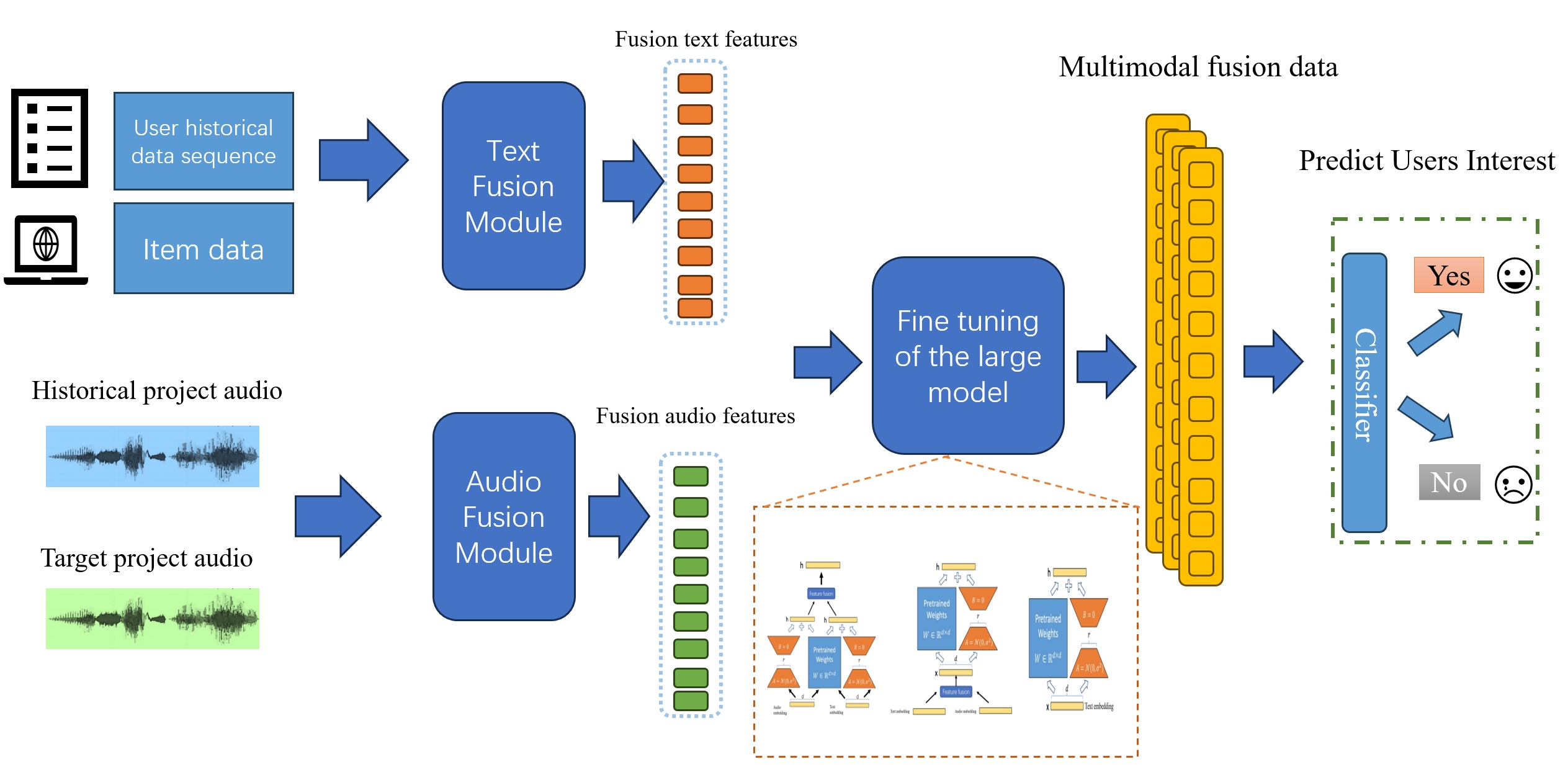}
    \caption{Model Framework}
    \label{fig:your_image_label}
\end{figure}

\subsection{Internal Structure of the Large Model recommender System}

Figure 2 illustrates the internal structure of the large model recommender system. Initially, the input data comprises audio and text vectors. Attention masks and position embedding features are then applied to the text vectors. Both audio and text vectors are subsequently fed into the large model for fine-tuning. The resulting audio and text features are fused through feature fusion. Finally, the fused features are input into a classifier for classification.
\begin{figure}[h]
    \centering
    \includegraphics[width=\linewidth]{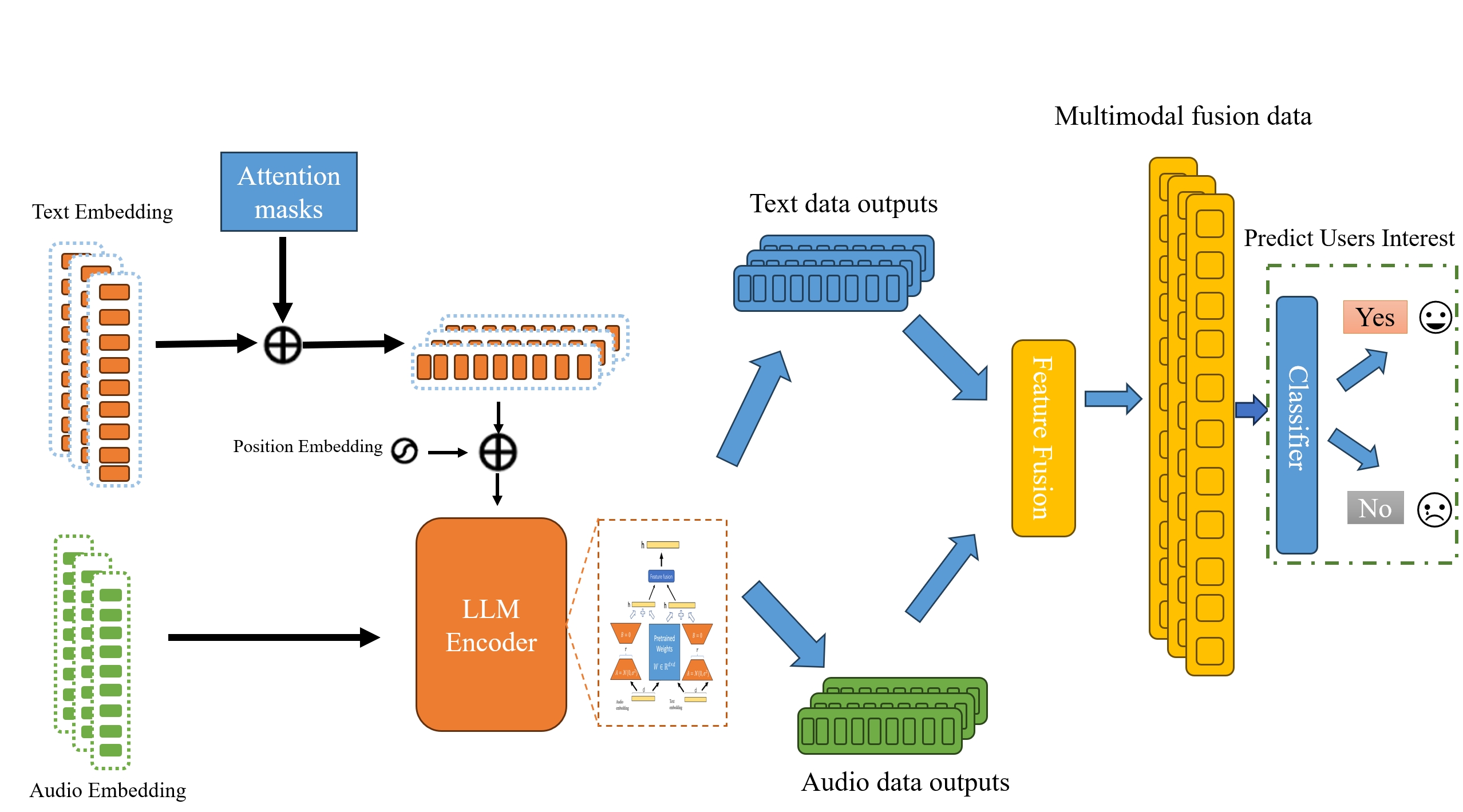}
    \caption{Internal Structure of the Large Model}
    \label{fig:your_image_label}
\end{figure}

\subsubsection{Audio Embedding Extraction}

To effectively extract time-frequency features from audio, this study uses a Mel filter bank(FBank) for audio feature representation\cite{li2021audio} as figure 3. Specifically, the Mel filter bank is computed with a frame length of 25 milliseconds and a frame shift of 10 milliseconds. Inspired by \cite{cao2022swin}, after feature processing, the audio embedding model processes these FBank features through a series of fully connected layers with nonlinear activation functions (SiLU). 

The specific structure of the audio embedding network is as follows: initially, the low-dimensional FBank input is mapped to a 256-dimensional hidden space through a linear transformation. This is followed by the application of the SiLU activation function to introduce nonlinearity, and further mapping to a higher-dimensional space to align with the text features dimensions, finally undergoing batch normalization to ensure stable gradient propagation and faster convergence.

\begin{figure}[h]
    \centering
    \includegraphics[width=\linewidth]{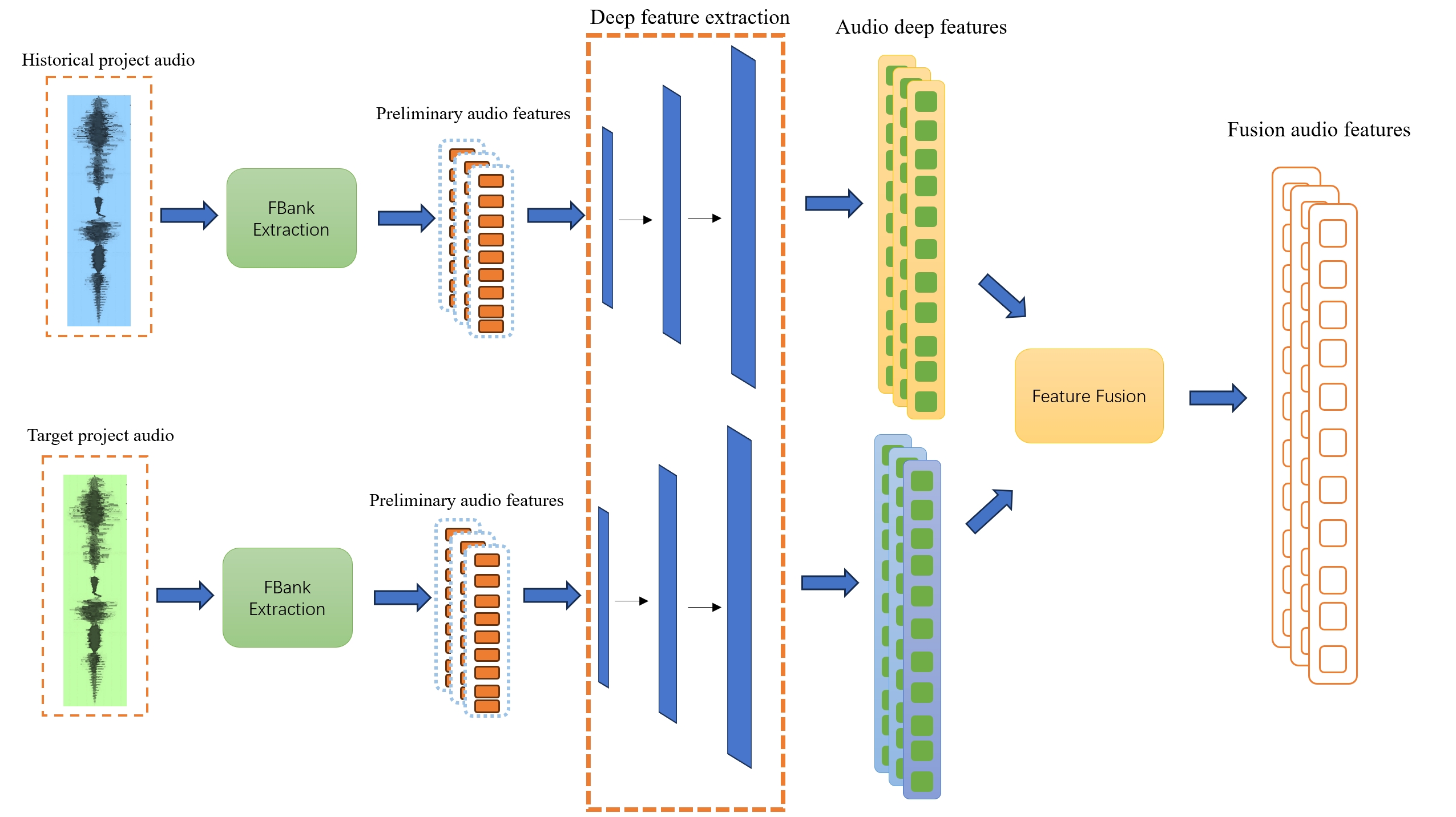}
    \caption{Audio Embedding Extraction}
    \label{fig:your_image_label}
\end{figure}
\subsubsection{Recommender system fine-tuning modules}

The features generated by the audio and text are used as fine-tuning data for various LoRA modules. This study utilizes the RoBERTa model \cite{delobelle2020robbert}. The causal self-attention parameters of the model are adjusted using parameter-efficient low-rank adaptation (LoRA), while all other parameters remain frozen.

To explore the impact of different LoRA module configurations on recommender system (RS) performance and optimize the integration of text and audio data, this paper adjusts the original LoRA fine-tuning module. As shown in Figure 4, Figure 4(a) illustrates a structure with two LoRA low-rank matrix modules, which independently fine-tune audio and text data before feature fusion through the Fusion module. Figure 4(b) features a single LoRA low-rank matrix module that first integrates audio and text data before performing overall fine-tuning. Figure 4(c) follows \cite{bao2023tallrec}, using a single LoRA low-rank matrix module to fine-tune text data. Through these different model design configurations, this study evaluates their impact on model performance.

In the ablation study, I examine the impact of different LoRA modules on the performance of the recommender system.

\begin{figure}[h]
    \centering
    \includegraphics[width=\linewidth]{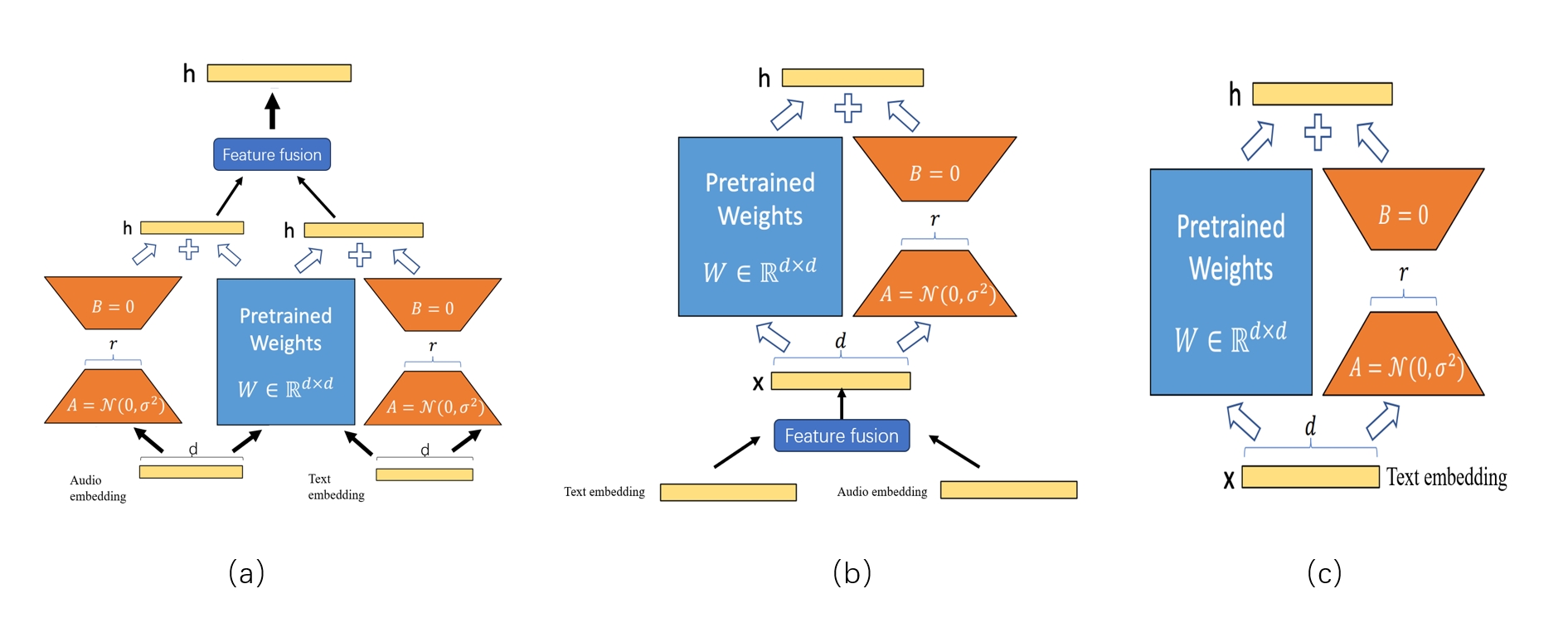}
    \caption{Three Fine-Tuning LORA Modules}
    \label{fig:your_image_label}
\end{figure}

\subsection{Instruction Tuning}

Instruction tuning is a core technique for training large-scale language models based on human-annotated instructions and responses\cite{zhang2023instruction}. This paper aims to perform recommendation tuning (rec-tuning) on large language models (LLMs) using recommendation data to develop a large-scale recommender system capable of predicting user preferences for new items. To achieve this, I convert recommendation data into the format required for instruction tuning, as illustrated in Table I.

\begin{table}[h]
    \centering
    \caption{A tuning sample for rec-tuning.}
    \label{fig:table_screenshot}
    \includegraphics[width=\linewidth]{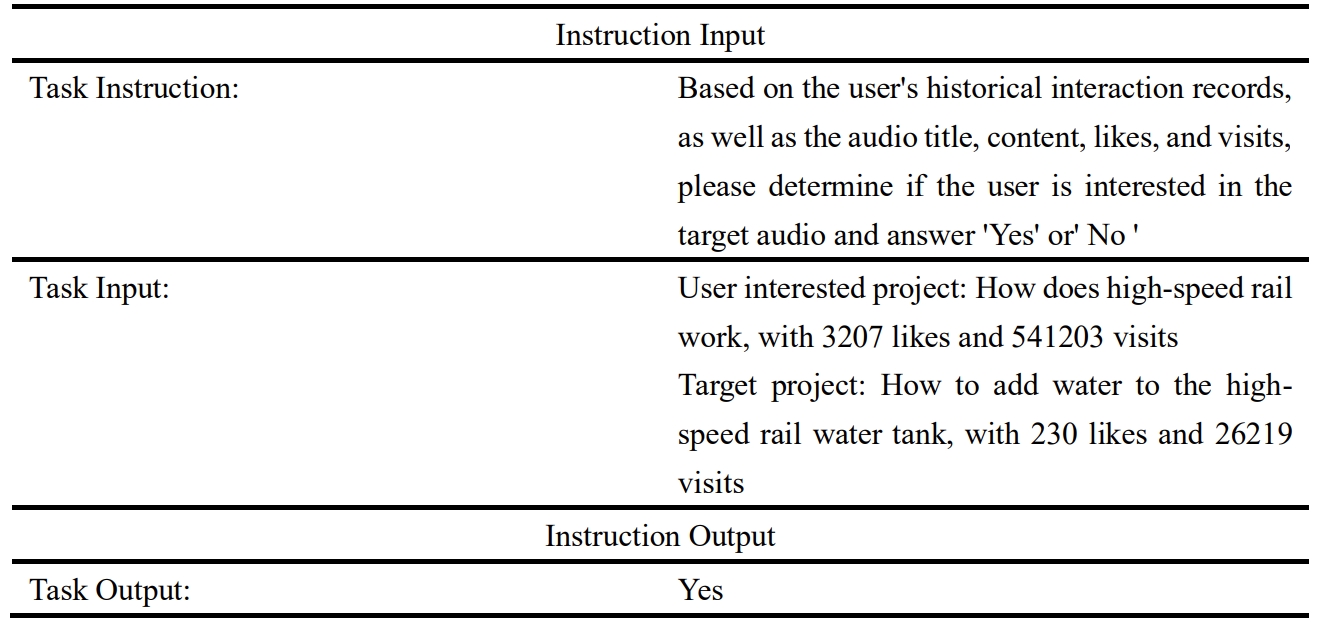}
\end{table}

\section{Experiment}

\subsubsection{Dataset Description}

MicroLens is a large-scale dataset focused on micro-video recommender systems \cite{ni2023content}. The MicroLens dataset contains 1 billion interaction records between users and videos, involving 34 million users and 1 million micro-videos. Each video provides comprehensive multimodal information, including titles, video content, user likes, and views. Audio data is extracted from the videos using an audio extraction program. Since critical information in short videos is typically found at the beginning \cite{lu2019fifteen}, to better handle the audio data features, the audio is truncated to the first 30 seconds.

To test whether the model can effectively address the cold-start problem of recommender systems with very limited training data, I use a few-shot training setting, where a limited number of samples are randomly selected from the training set for model training, as followed by \cite{bao2023tallrec}. This is referred to as the "K-shot" training setting, where K represents the number of training samples used. By setting a very small K value, it is possible to evaluate whether the method can quickly gain recommendation capability from the LLM with very limited training data.

\subsubsection{Comparison Models}
Since this approach utilizes historical interactions to predict subsequent interactions, similar to sequential recommendation methods, the following sequential models are compared: GRU4Rec\cite{hidasi2015session} is a recommendation algorithm based on Gated Recurrent Units. It captures the dynamic interests of users and makes recommendations based on historical behavior sequences. SASRec\cite{kang2018self} is a sequence recommendation model based on the self-attention mechanism, which uses self-attention layers from the Transformer to model dependencies in user behavior sequences. GCN\cite{he2020lightgcn} is a graph neural network model designed to learn node feature representations by performing convolution operations on graph-structured data. GraphSAGE\cite{hamilton2017inductive} is a graph neural network model focused on handling large-scale graph data, which is particularly useful for managing large-scale user-item interaction graphs.

\subsubsection{Evaluation Metrics}

Since this model aims to predict user preferences for a given target item as a binary classification problem, this paper uses the widely adopted evaluation metric in recommendation systems: the Area Under the Receiver Operating Characteristic Curve (AUC)\cite{chicco2023matthews}.

\subsubsection{Implementation Details}

To ensure consistent sequence length, this paper standardizes the text length to 512, which matches the model's maximum supported length.

For training all LoRA modules, binary cross-entropy loss is used, with default LoRA hyperparameters set to R = 4. The optimizer is Adam. To enhance model convergence speed and avoid instability during training, a two-phase learning rate scheduling scheme is employed. Initially, a warm-up scheduler is used to gradually increase the learning rate; subsequently, a step decay learning rate is applied during the main training phase. In the warm-up phase, the learning rate increases from a minimal value ($10^{-9}$) to the set initial learning rate over 50 iterations. After the warm-up phase, the learning rate begins to gradually decay, linearly decreasing from the initial value to 0. The batch size is set to 8, with gradient accumulation performed every 4 iterations. For each training epoch, incomplete batches of data are discarded. The training process is completed on a NVIDIA RTX 3080 16GB GPU.

\subsection{Performance Comparison}

As shown in Table II, this study evaluates various recommendation models, including GRU4REC, SASRec, GCN, GraphSAGE, and ATFLRec, under few-shot learning settings with sample sizes of 100 and 500. The results reveal that ATFLRec performs exceptionally well in both few-shot settings, achieving AUC values of 0.6042 and 0.6708, respectively.

The superior performance of ATFLRec can be attributed to its advanced integration of text and audio modalities, which enhances the model's ability to capture user preferences. The higher AUC values indicate that ATFLRec is more effective at distinguishing between positive and negative recommendations, thus offering more personalized and accurate suggestions.

In contrast, traditional deep learning recommendation methods consistently perform poorly in the few-shot training settings. This suggests that conventional deep learning approaches require a large number of training parameters and struggle to rapidly acquire recommendation capabilities with limited training samples.

Furthermore, for graph neural network-based recommendation systems, GraphSAGE demonstrates its strength through its neighbor sampling and aggregation mechanisms, which enable the model to maintain high computational efficiency when processing large-scale graph data. However, this sampling mechanism has limitations, as it may overlook significant graph structural information, potentially affecting the final recommendation performance\cite{yang2020understanding}.
\begin{table}[h]
    \centering
    \caption{Performance Comparison}
    \label{fig:table_screenshot}
    \includegraphics[width=\linewidth]{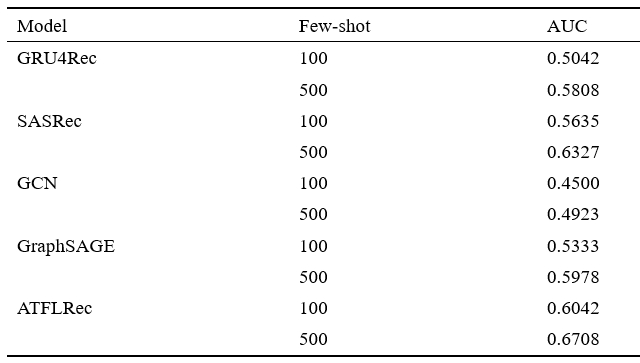}
\end{table}

\subsection{Ablation Study: Assessing the Impact of Fine-Tuning Modules on Recommender System Performance}

The purpose of this section is to investigate the impact of different LoRA fine-tuning modules on the performance of recommender systems. As shown in Table III, the analysis reveals that using separate LoRA modules to train text and audio data, followed by modal fusion for recommendation, yields the best performance. Conversely, models employing a single LoRA module and considering only textual information exhibit the worst performance. This suggests that integrating audio data with text data can enhance the performance of recommender systems.

Moreover, the results indicate that training audio and text data separately with different low-rank matrices for subsequent fusion produces better performance than using a single low-rank matrix to handle both modalities simultaneously. This finding provides valuable insights into the fine-tuning of multimodal data.
\begin{table}[h]
    \centering
    \caption{Ablation Study}
    \label{fig:table_screenshot}
    \includegraphics[width=\linewidth]{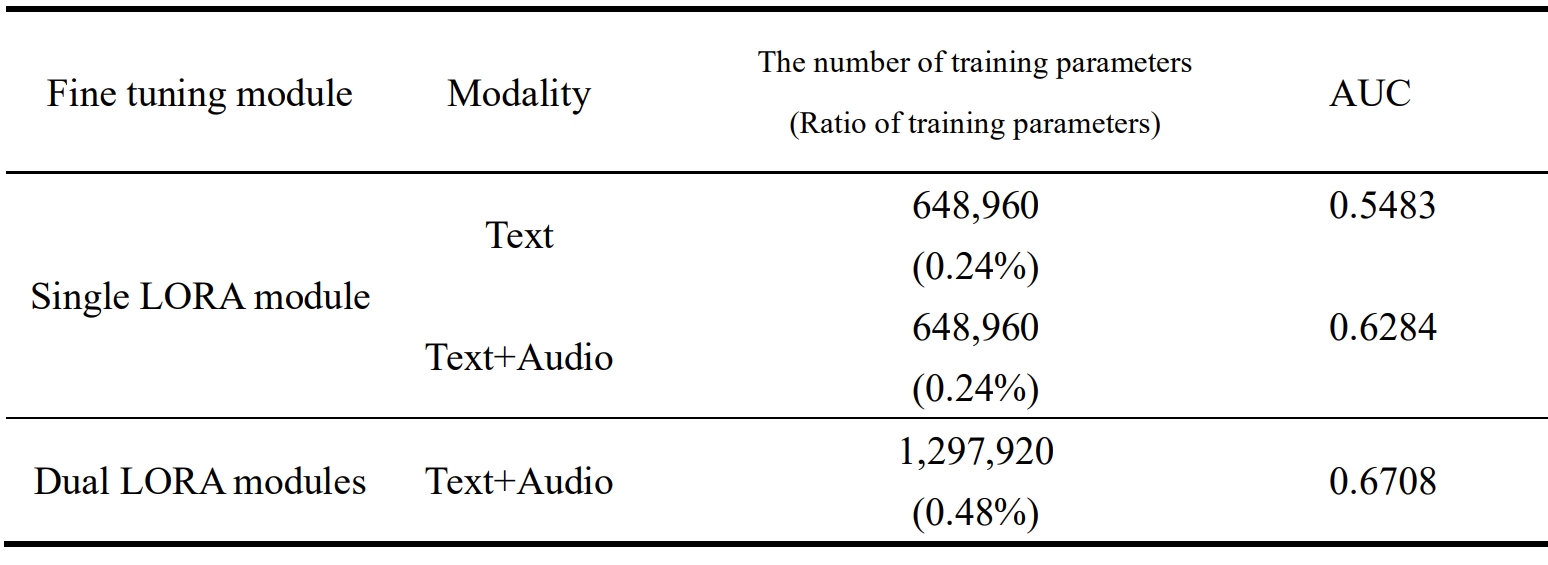}
\end{table}

\subsection{Exploring the Impact of Feature Fusion from Same and Different Modalities on Recommender System Performance}
This section explores the impact of data fusion methods on system performance. Modal fusion can be categorized into fusion within the same modality and fusion across different modalities. During the fusion process, some important features of the data are strengthened and amplified, while less important features may be omitted. To identify the optimal fusion method, this paper conducted experiments on both audio-only fusion and audio-text fusion, as shown in Table IV. The results indicate that the recommender system performs optimally with the maximum pooling operation. In the final step of the model, audio and text features extracted from LoRA training are stacked and fused using pooling methods. Among these, the Sum pooling operation yields the best performance for the recommender system\cite{almeida2022complementarity}.
\begin{table}[h]
    \centering
    \caption{Impact of Feature Fusion from Same and Different Modalities}
    \label{fig:table_screenshot}
    \includegraphics[width=\linewidth]{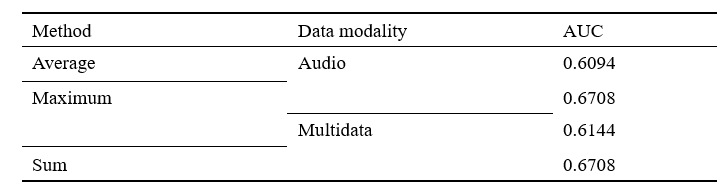}
\end{table}

\subsection{Exploring the Impact of the Number of FBanks}

As shown in Table V, FBank captures the energy distribution of audio signals across various frequency bands, helping the model better understand frequency characteristics\cite{abdul2022mel}. Therefore, the choice of the number of FBanks used can significantly impact the final performance of the recommender system. The system achieves the best performance with 80 FBanks.
\begin{table}[h]
    \centering
    \caption{Impact of the Number of FBanks}
    \label{fig:table_screenshot}
    \includegraphics[width=\linewidth]{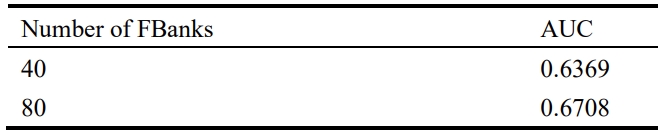}
\end{table}
\section{Conclusion}

This paper introduces ATFLRec, an innovative multimodal recommender system that combines audio and text data to enhance performance. Integrating these modalities into a large-scale language model allows for more precise capture of user preferences and more personalized recommendations. The system's core advantage is its efficient use of low-rank adaptation (LoRA) modules, which improve recommendation capability while saving computational resources. Experimental results show that ATFLRec outperforms traditional deep learning methods in few-shot settings, with the best performance achieved when audio and text data are separately trained and then fused using different LoRA modules. Additionally, maximum pooling of audio and text features and a configuration of 80 FBANKs optimize performance, capturing frequency characteristics effectively. Future research will focus on exploring interactions between modalities and optimizing techniques for multimodal data fusion.

\bibliographystyle{unsrt}
\bibliography{references}

\begin{thebibliography}{10}

\bibitem{tongxiao2011value}
Tongxiao~(Catherine) Zhang, Ritu Agarwal, and Henry~C Lucas~Jr.
\newblock The value of it-enabled retailer learning: personalized product recommendations and customer store loyalty in electronic markets.
\newblock {\em MIS quarterly}, pages 859--881, 2011.

\bibitem{singh2021recommender}
Pradeep~Kumar Singh, Pijush Kanti~Dutta Pramanik, Avick~Kumar Dey, and Prasenjit Choudhury.
\newblock Recommender systems: an overview, research trends, and future directions.
\newblock {\em International Journal of Business and Systems Research}, 15(1):14--52, 2021.

\bibitem{deldjoo2021multimedia}
Yashar Deldjoo, Markus Schedl, Bal{\'a}zs Hidasi, Yinwei Wei, and Xiangnan He.
\newblock Multimedia recommender systems: Algorithms and challenges.
\newblock In {\em Recommender systems handbook}, pages 973--1014. Springer, 2021.

\bibitem{afchar2022explainability}
Darius Afchar, Alessandro Melchiorre, Markus Schedl, Romain Hennequin, Elena Epure, and Manuel Moussallam.
\newblock Explainability in music recommender systems.
\newblock {\em AI Magazine}, 43(2):190--208, 2022.

\bibitem{jannach2017recurrent}
Dietmar Jannach and Malte Ludewig.
\newblock When recurrent neural networks meet the neighborhood for session-based recommendation.
\newblock In {\em Proceedings of the eleventh ACM conference on recommender systems}, pages 306--310, 2017.

\bibitem{wu2022graph}
Shiwen Wu, Fei Sun, Wentao Zhang, Xu~Xie, and Bin Cui.
\newblock Graph neural networks in recommender systems: a survey.
\newblock {\em ACM Computing Surveys}, 55(5):1--37, 2022.

\bibitem{silva2019pure}
N{\'\i}collas Silva, Diego Carvalho, Adriano~CM Pereira, Fernando Mour{\~a}o, and Leonardo Rocha.
\newblock The pure cold-start problem: A deep study about how to conquer first-time users in recommendations domains.
\newblock {\em Information Systems}, 80:1--12, 2019.

\bibitem{zhao2024recommender}
Zihuai Zhao, Wenqi Fan, Jiatong Li, Yunqing Liu, Xiaowei Mei, Yiqi Wang, Zhen Wen, Fei Wang, Xiangyu Zhao, Jiliang Tang, et~al.
\newblock Recommender systems in the era of large language models (llms).
\newblock {\em IEEE Transactions on Knowledge and Data Engineering}, 2024.

\bibitem{hu2021lora}
Edward~J Hu, Yelong Shen, Phillip Wallis, Zeyuan Allen-Zhu, Yuanzhi Li, Shean Wang, Lu~Wang, and Weizhu Chen.
\newblock Lora: Low-rank adaptation of large language models.
\newblock {\em arXiv preprint arXiv:2106.09685}, 2021.

\bibitem{zhou2024survey}
Zixuan Zhou, Xuefei Ning, Ke~Hong, Tianyu Fu, Jiaming Xu, Shiyao Li, Yuming Lou, Luning Wang, Zhihang Yuan, Xiuhong Li, et~al.
\newblock A survey on efficient inference for large language models.
\newblock {\em arXiv preprint arXiv:2404.14294}, 2024.

\bibitem{bao2023tallrec}
Keqin Bao, Jizhi Zhang, Yang Zhang, Wenjie Wang, Fuli Feng, and Xiangnan He.
\newblock Tallrec: An effective and efficient tuning framework to align large language model with recommendation.
\newblock In {\em Proceedings of the 17th ACM Conference on Recommender Systems}, pages 1007--1014, 2023.

\bibitem{cui2022m6}
Zeyu Cui, Jianxin Ma, Chang Zhou, Jingren Zhou, and Hongxia Yang.
\newblock M6-rec: Generative pretrained language models are open-ended recommender systems.
\newblock {\em arXiv preprint arXiv:2205.08084}, 2022.

\bibitem{fathullah2024prompting}
Yassir Fathullah, Chunyang Wu, Egor Lakomkin, Junteng Jia, Yuan Shangguan, Ke~Li, Jinxi Guo, Wenhan Xiong, Jay Mahadeokar, Ozlem Kalinli, et~al.
\newblock Prompting large language models with speech recognition abilities.
\newblock In {\em ICASSP 2024-2024 IEEE International Conference on Acoustics, Speech and Signal Processing (ICASSP)}, pages 13351--13355. IEEE, 2024.

\bibitem{truong2021multi}
Quoc-Tuan Truong, Aghiles Salah, and Hady Lauw.
\newblock Multi-modal recommender systems: Hands-on exploration.
\newblock In {\em Proceedings of the 15th ACM Conference on Recommender Systems}, pages 834--837, 2021.

\bibitem{li2021audio}
Zhanli Li and Pengfei Song.
\newblock Audio similarity detection algorithm based on siamese lstm network.
\newblock In {\em 2021 6th International Conference on Intelligent Computing and Signal Processing (ICSP)}, pages 182--186. IEEE, 2021.

\bibitem{cao2022swin}
Hu~Cao, Yueyue Wang, Joy Chen, Dongsheng Jiang, Xiaopeng Zhang, Qi~Tian, and Manning Wang.
\newblock Swin-unet: Unet-like pure transformer for medical image segmentation.
\newblock In {\em European conference on computer vision}, pages 205--218. Springer, 2022.

\bibitem{delobelle2020robbert}
Pieter Delobelle, Thomas Winters, and Bettina Berendt.
\newblock Robbert: a dutch roberta-based language model.
\newblock {\em arXiv preprint arXiv:2001.06286}, 2020.

\bibitem{zhang2023instruction}
Shengyu Zhang, Linfeng Dong, Xiaoya Li, Sen Zhang, Xiaofei Sun, Shuhe Wang, Jiwei Li, Runyi Hu, Tianwei Zhang, Fei Wu, et~al.
\newblock Instruction tuning for large language models: A survey.
\newblock {\em arXiv preprint arXiv:2308.10792}, 2023.

\bibitem{ni2023content}
Yongxin Ni, Yu~Cheng, Xiangyan Liu, Junchen Fu, Youhua Li, Xiangnan He, Yongfeng Zhang, and Fajie Yuan.
\newblock A content-driven micro-video recommendation dataset at scale.
\newblock {\em arXiv preprint arXiv:2309.15379}, 2023.

\bibitem{lu2019fifteen}
Xing Lu and Zhicong Lu.
\newblock Fifteen seconds of fame: A qualitative study of douyin, a short video sharing mobile application in china.
\newblock In {\em Social Computing and Social Media. Design, Human Behavior and Analytics: 11th International Conference, SCSM 2019, Held as Part of the 21st HCI International Conference, HCII 2019, Orlando, FL, USA, July 26-31, 2019, Proceedings, Part I 21}, pages 233--244. Springer, 2019.

\bibitem{hidasi2015session}
B~Hidasi.
\newblock Session-based recommendations with recurrent neural networks.
\newblock {\em arXiv preprint arXiv:1511.06939}, 2015.

\bibitem{kang2018self}
Wang-Cheng Kang and Julian McAuley.
\newblock Self-attentive sequential recommendation.
\newblock In {\em 2018 IEEE international conference on data mining (ICDM)}, pages 197--206. IEEE, 2018.

\bibitem{he2020lightgcn}
Xiangnan He, Kuan Deng, Xiang Wang, Yan Li, Yongdong Zhang, and Meng Wang.
\newblock Lightgcn: Simplifying and powering graph convolution network for recommendation.
\newblock In {\em Proceedings of the 43rd International ACM SIGIR conference on research and development in Information Retrieval}, pages 639--648, 2020.

\bibitem{hamilton2017inductive}
Will Hamilton, Zhitao Ying, and Jure Leskovec.
\newblock Inductive representation learning on large graphs.
\newblock {\em Advances in neural information processing systems}, 30, 2017.

\bibitem{chicco2023matthews}
Davide Chicco and Giuseppe Jurman.
\newblock The matthews correlation coefficient (mcc) should replace the roc auc as the standard metric for assessing binary classification.
\newblock {\em BioData Mining}, 16(1):4, 2023.

\bibitem{yang2020understanding}
Zhen Yang, Ming Ding, Chang Zhou, Hongxia Yang, Jingren Zhou, and Jie Tang.
\newblock Understanding negative sampling in graph representation learning.
\newblock In {\em Proceedings of the 26th ACM SIGKDD international conference on knowledge discovery \& data mining}, pages 1666--1676, 2020.

\bibitem{almeida2022complementarity}
Adolfo Almeida, Johan~Pieter de~Villiers, Allan De~Freitas, and Mergandran Velayudan.
\newblock The complementarity of a diverse range of deep learning features extracted from video content for video recommendation.
\newblock {\em Expert Systems with Applications}, 192:116335, 2022.

\bibitem{abdul2022mel}
Zrar~Kh Abdul and Abdulbasit~K Al-Talabani.
\newblock Mel frequency cepstral coefficient and its applications: A review.
\newblock {\em IEEE Access}, 10:122136--122158, 2022.

\end{thebibliography}
\vspace{12pt}
\color{red}

\end{document}